\documentclass[11pt,a4paper]{scrartcl}

\RequirePackage[
  pdfstartview=FitH,
  breaklinks=true,
  bookmarks=true,
  colorlinks=true,
  linkcolor= blue,
  anchorcolor=blue,
  citecolor=blue,
  filecolor=blue,
  menucolor=blue,
  urlcolor=blue
  ]{hyperref}
  \AtBeginDocument{
  \hypersetup{
    pdfauthor = {Antonin Schrab},
    colorlinks = true,
    urlcolor = cyan,
    linkcolor = blue,
    citecolor = orange,
    pdftitle = {Title - compilation : \today}
  }
}

\usepackage[utf8]{inputenc}
\usepackage[T1]{fontenc}
\usepackage{url}
\usepackage{booktabs}
\usepackage{amsfonts}
\usepackage{nicefrac}
\usepackage{microtype}
\usepackage{xcolor}
\usepackage{mathtools}
\usepackage{amsmath}
\usepackage{amsthm}
\usepackage{tcolorbox}
\usepackage{placeins}
\usepackage{enumitem}

\usepackage[round]{natbib}

\usepackage[top=3cm, bottom=3cm, left=3cm, right=3cm]{geometry}
\usepackage{color}
\usepackage{amsmath}
\usepackage{amssymb}
\usepackage{amsfonts}
\usepackage{amsthm}
\usepackage{tcolorbox}
\usepackage[english]{babel}
\usepackage{fancyhdr}
\usepackage[mathcal]{eucal}
\usepackage{cleveref}

\fancypagestyle{firststyle}
{
   \fancyhf{}
   
   \cfoot{\thepage}
}

\crefname{assumption}{Assumption}{Assumptions}
\crefname{equation}{Equation}{Equations}
\crefname{figure}{Figure}{Figures}
\crefname{table}{Table}{Tables}
\crefname{section}{Section}{Sections}
\crefname{theorem}{Theorem}{Theorems}
\crefname{lemma}{Lemma}{Lemmas}
\crefname{corollary}{Corollary}{Corollaries}
\crefname{example}{Example}{Examples}
\crefname{appendix}{Appendix}{Appendices}
\crefname{remark}{Remark}{Remarks}

\theoremstyle{plain}

\DeclareMathAlphabet{\mathcal}{OMS}{cmsy}{m}{n}

\begin{document}

\thispagestyle{firststyle}
{\center
{\huge 
\textbf{\textsc{Discussion of}}
\\
\textbf{\textsc{`Multiscale Fisher's Independence Test}}
\\
\bigskip
\textbf{\textsc{for Multivariate Dependence'}}
} \\

\bigskip
\bigskip

{\fontsize{12}{15}\selectfont 
\textbf{\textsc{Antonin Schrab}} \\
Centre for Artificial Intelligence\\
Gatsby Computational Neuroscience Unit\\
University College London and Inria London \\
\texttt{a.schrab@ucl.ac.uk} \\
\bigskip
\textbf{\textsc{Wittawat Jitkrittum}} \\
Google Research\\
New York \\ 
\bigskip
\textbf{\textsc{Zolt{\'a}n Szab{\'o}}} \\
Department of Statistics \\
London School of Economics and Political Science \\
\bigskip
\textbf{\textsc{Dino Sejdinovic}} \\
Department of Statistics \\
University of Oxford \\
\bigskip
\textbf{\textsc{Arthur Gretton}} \\
Gatsby Computational Neuroscience Unit\\
University College London \\

\bigskip
\bigskip

\today

\bigskip
\bigskip
}
}

\centerline{\Large{\textbf{\textsf{Abstract}}}}

\begin{abstract}
We discuss how MultiFIT, the Multiscale Fisher's Independence Test for Multivariate Dependence proposed by \citet{gorsky2018multiscale}, compares to existing linear-time kernel tests based on the Hilbert-Schmidt independence criterion (HSIC).
We highlight the fact that the levels of the kernel tests at any finite sample size can be controlled exactly, as it is the case with the level of MultiFIT.
In our experiments, we observe some of the performance limitations of MultiFIT in terms of test power.
\end{abstract}

\section{Introduction}
We read with  interest the work of \citet{gorsky2018multiscale} on statistical dependence testing using a Multiscale Fisher's Independence Test (MultiFIT). 
The procedure consists in first transforming the data to map to the unit ball, then performing univariate Fisher's exact tests of independence on a collection of $2 \times 2$ contingency tables, and finally correcting for the use of multiple testing.
The collection is obtained using a divide-and-conquer approach with a coarse-to-fine procedure: 
the unit ball is partitioned into cuboids and $2 \times 2$ contingency tables of counts of samples in the cuboids are tested, the cuboids with small associated $p$-values are then further partitioned at finer resolutions and tested again, etc. 
This approach has a number of advantages, chief among them that the test is multivariate, the computational cost is in general $\mathcal{O}(n\log n)$ as a function of sample size $n$, and the test threshold is exact at any sample size (not an asymptotic limit).

The problem of computationally efficient, linear-time dependence testing is important to address, and a number of approaches have been proposed in the machine learning and statistics literature.
In the present discussion, we will provide brief descriptions of certain of these approaches, enumerating their advantages and disadvantages in comparison with MultiFIT. 
We will evaluate the performance in terms of power of all tests  on both synthetic and real-world data.

We begin by placing the statistics against which the authors compared, namely the (Brownian) distance covariance  and its generalisations \citep{szekely2009,lyons2013}, within the broader framework of kernel-based independence testing.
\citet[Theorem 24]{SejSriGreFuk2013} established that the (generalised) distance covariance is an instance of a Hilbert-Schmidt independence criterion \citep[HSIC;][]{GreBouSmoSch2005}, which is the Hilbert-Schmidt norm of a covariance operator between features of $X$ and of $Y$ in respective reproducing kernel Hilbert spaces (RKHSs). 
When these reproducing kernel Hilbert spaces are sufficiently rich (i.e., characteristic, \citealp{SriGreFukSchetal10}), then HSIC is zero if and only if $X$ and $Y$ are independent \citep{Gretton15,SzaSri18}. 
This is always the case for exponentiated quadratic kernels, and is also true for the family of distance-induced kernels that define the (Brownian) distance covariance, subject to appropriate moment conditions \citep[Proposition 29, Remark 31]{SejSriGreFuk2013}.
Statistical tests of independence using HSIC have been proposed by \cite{GreFukTeoSonSchSmo08,ChwGre14,ChwSejGre14}, with computational cost $\mathcal{O}(n^2)$. 
In the next section, we will describe approaches developed from these kernel statistics to yield greater computational efficiency and improved power.

As with all kernel methods, the power of the distance covariance in statistical testing can be improved by a suitable selection of parameters \citep{li2019optimality}. 
In the case of the distance covariance on $\mathbb{R}^d$, this amounts to raising the Euclidean distances between samples to a power $a\in(0,2],$ as discussed by \citet[Section 8.2]{SejSriGreFuk2013}.
When exponentiated quadratic kernels are used, \citet{albert2019adaptive} propose an adaptive minimax quadratic-time test for alternatives over Sobolev balls, which aggregates tests with varying kernel parameters. 
\citet{schrab2022efficient} have recently proposed a linear-time variant of this adaptive test and have quantified the cost incurred in the minimax rate for computational efficiency.
These approaches provide a systematic instantiation of the notions corresponding to ``departures from independence at coarse-to-fine scales'' as discussed by \citeauthor{gorsky2018multiscale}.

\section{Linear-time HSIC tests}
\label{sec:kernel_tests}

We note that a number of methodological contributions pertaining to large-scale versions of tests based on HSIC and related quantities have appeared in the prior literature.
Large-scale approximations to kernel methods are a well-studied field of research: among the most widely used approximation paradigms are the \emph{Nystr\"om approximation} \citep{williams01using}, where the corresponding RKHS is approximated by a subspace spanned by the so-called inducing or landmark points, and the \emph{Random Fourier Features approximation} (RFF; \citealt{rahimi07random}), where explicit feature maps are constructed via Fourier representations of shift-invariant kernels.
\citet{zhang2018large} employ both of these approaches to construct computationally efficient HSIC-based independence tests, demonstrating significant savings in computation time and memory. 
With those approximations, the asymptotic null distribution is estimated in linear time using an eigendecomposition of primal covariance matrices, which is computationally more efficient than using permutations or sampling directly from the null distribution.
We refer to the two resulting tests as NyHSIC (Nystr\"om approximation, with cost $\mathcal{O}(m_x m_y n)$, where $m_x, m_y$ are the numbers of Nystr\"om inducing points for the respective feature spaces) and FoHSIC (random Fourier feature approximation, with cost $\mathcal{O}(d_x d_y n)$, where $d_x, d_y$ are the numbers of Fourier features used to approximate the respective feature spaces).
We remark that, for the tests to be consistent, the number of Nystr\"om points/Fourier features must grow with increasing sample size $n:$ this is analogous to the partition refinement of  MultiFIT with $n$ \citep[Theorem 2.3]{gorsky2018multiscale}.

The Finite Set Independence Criterion (FSIC; \citealt{jitkrittum2017adaptive}) is an adaptive linear-time independence test that is applicable to high-dimensional problems.
Briefly, the FSIC statistic is defined as the average of covariances of a finite number of real analytic functions (i.e., features) defined on the joint domain of the two multivariate variables in consideration.
While the use of finitely many analytic functions resembles the RFF-based HSIC test discussed above, FSIC's features are {\em adaptive}, in that they are chosen to maximize a lower bound on the test power (on a held-out sample, which reduces the number of samples available for testing, but nonetheless results in a net improvement in test power: see experiments for details).
Under smoothness conditions on the kernels, the FSIC test is consistent for any finite number of features used (see Proposition 2 of \citealt{jitkrittum2017adaptive}). 
In our experiments, we consider the permutation-based test which uses the Normalized FSIC (NFSIC).
The normalized variant NFSIC has a distribution-free asymptotic null distribution (chi-squared), further facilitating fast testing on a large-scale dataset by avoiding permutations.
Note that both feature optimization for a local optimum, and the statistical test using NFSIC, can be accomplished in linear time.

\vspace{-0.14cm}
\section{Exact control of non-asymptotic level}
\vspace{-0.1cm}

An important theoretical result of \citet[Corollary 2.1]{gorsky2018multiscale} is that MultiFIT attains exact control of the level at any finite sample size: this means that the test is always well-calibrated.
We emphasize that  control of the non-asymptotic  level is also guaranteed for kernel tests using permutations, and that this control is {\em exact}.
\citet[Proposition 1]{albert2019adaptive} prove that the quadratic-time HSIC test, which uses permutations for quantile estimation, exactly controls  the non-asymptotic level. 
The proof relies only on the exchangeability under the null of the original test statistic with the permuted test statistics, and the exact control of the level of a permutation-based test can more generally be guaranteed whenever permutation-invariance holds (including for our linear-time kernel tests), as explained by \citet[Section 2.1]{kim2020minimax}.
We also point out the work of \citet[Sections 2.1 and 3.1]{gretton2010consistent} who construct multivariate nonparametric tests of independence based on the $L_1$ and KL divergences computed on partitioned spaces, with distribution-free thresholds based on finite sample bounds.
These thresholds exactly control the non-asymptotic level at any sample size, however, they are shown to be conservative in comparison with asymptotic thresholds for the same statistics.
Finally, \citet[Sections 5 \& 7]{kim2020minimax} also propose a binning-based independence test: a permutation-based multinomial test is performed on the discretized data.
Their test exactly controls the non-asymptotic level; it is also shown to be minimax adaptive and optimal over the H{\"o}lder class of density functions.

\section{Experiments}

In our experiments, we compare the performance in terms of test power for MultiFIT and for the linear-time HSIC tests presented in Section~\ref{sec:kernel_tests} \citep[\href{https://example.com}{code};][]{jitkrittum2017adaptive}.
We ran MultiFIT with the resolution-specific approach to multiple testing, with Holm's method on $p$-values with mid-$p$ correction.
This method allows for  early stopping when sufficient evidence for rejecting the null has been observed in the first few resolutions, and does not impact the quality of the test.
At first we ran the MultiFIT test with its default parameter $R^\star=1$; this parameter sets the resolution until which all cuboids are necessarily tested regardless of the test results, after which only cuboids where dependence has been detected are partitioned further to be tested at the next resolution.
The parameter $R^\star=1$ means that all cuboids in the first two resolutions (0 and 1) are necessarily considered.
Having observed low test power in certain cases for MultiFIT with this default parameter, we  increased the value to $R^\star=2$. This increases the power, but can entail a significant computational cost in high-dimensional settings.
This parameter tuning needs to be done a posteriori (after having observed low power), which is a limitation of the MultiFIT method.

\begin{figure}[b]
\includegraphics[width=\textwidth]{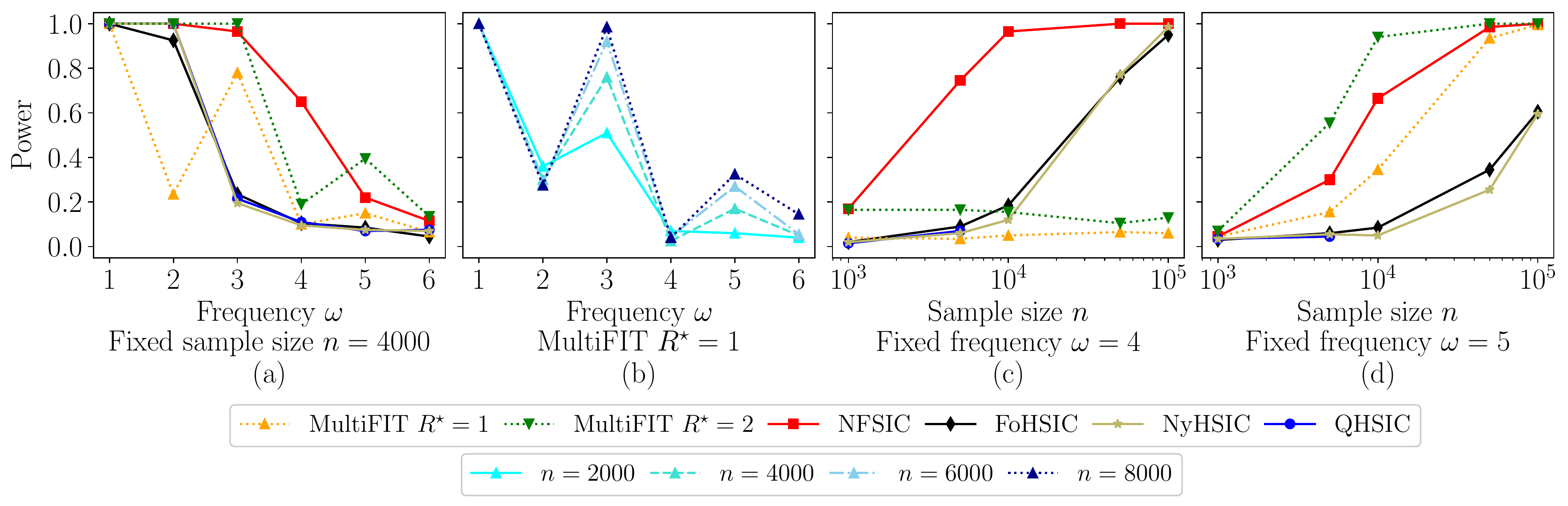}
\caption{Sinusoid experiment with joint density $p_{xy}\propto 1+\sin(\omega x)\sin(\omega y)$ on $(-\pi,\pi)^2$ for some frequency $\omega>0$.}
\label{fig:sin}
\end{figure}

For the kernel tests introduced in Section~\ref{sec:kernel_tests} with Gaussian kernel, we used:
NyHSIC with 10 randomly chosen inducing points,
FoHSIC with 10 random Fourier features,
{NFSIC} with 10 selected adaptive features,
and QHSIC which is the original quadratic-time HSIC test proposed by \citet{GreBouSmoSch2005}.
For NyHSIC and FoHSIC, the null is simulated by sampling 2000 points from the estimated asymptotic distribution.
For NFSIC and QHSIC, we use 500 permutations.
The adaptive NFSIC test requires a held-out sample for feature adaptation: thus, we used $n/2$ of the available samples for this purpose, and tested on the remaining $n/2$ samples (in other words, NFSIC tested on half the samples of the remaining tests). 
All tests have well-calibrated levels \citep{jitkrittum2017adaptive}.
The run times of the kernel tests vary depending on the parameter choices.

We reproduce three experiments proposed by \citet{jitkrittum2017adaptive}, namely the Sinusoid, Gaussian Sign and Million Song Dataset experiments.
To evaluate the test power, we repeat each experiment 200 times and plot the averages. 
For time complexity, we average the run times over 10 runs performed on an AMD Ryzen Threadripper 3960X 24 Cores 128Gb RAM CPU at 3.8GHz.

In \Cref{fig:sin}, we consider the Sinusoid problem where the variables $(X,Y)$ have joint density $p_{xy}\propto 1+\sin(\omega x)\sin(\omega y)$ on $(-\pi,\pi)^2$ for some frequency $\omega>0$. 
For sample size $n=4000$, we observe in \Cref{fig:sin}~(a) that the power of all kernel tests decreases as the frequency $w$ increases, which is expected since the departure from the null occurs at higher frequencies, becoming harder to detect. 
For MultiFIT $R^\star=1$, we observe that the test has lower power for the frequencies 2 and 4 than for the frequencies 3 and 5, which is at first surprising.
The same pattern is observed for MultiFIT $R^\star=2$ with higher power for frequency 5 than for frequency 4.
We see in \Cref{fig:sin}~(b) that, for MultiFIT $R^\star=1$ and $\omega=4$, the test power remains close to zero as the sample size increases, and that for frequency $\omega=2$, increasing the sample size actually decreases test power.
This disparity is confirmed in Figures~\ref{fig:sin}~(b)~\&~(c) where for $\omega=4$ increasing the sample size does not increase test power for MultiFIT with $R^\star\in\{1,2\}$, while it does for frequency 5.
For $\omega=5$, we even observe that MultiFIT $R^\star=2$ outperforms all other tests, however, it has almost zero power for $\omega=4$.
The reason for the low power of MultiFIT $R^\star=1$ at $\omega\in\{2,4\}$ is that the Fisher tests fail to detect dependence in coarse resolution, and so, finer resolutions are not considered.
For even frequencies $\omega$, the quadrants of $(-\pi,\pi)^2$ contain exactly full oscillations, and so the number of samples in the cuboid drawn uniformly and drawn from the joint $p_{xy}$ have the same distribution. 
Hence, MultiFIT $R^\star=1$ cannot detect dependence at any sample sizes for even frequencies.
Increasing $R^\star$ to $2$ only shifts the problem to higher resolutions/frequencies, as observed with $\omega = 4$ for MultiFIT $R^\star=2$.
In fact, for any choice of $R^\star\in\mathbb{N}\setminus\{0\}$, MultiFIT cannot detect dependence of the Sinusoid problem with frequency $\omega = 2^{R^\star}$ at any sample size (the same holds for the $\cos$ function).
This toy experiment reveals a more fundamental limitation of MultiFIT: the test will be `blind' to certain frequencies in the characteristic function of the joint density, and will not accumulate evidence at these frequencies in deciding whether to reject the null.

\begin{figure}[b]
\includegraphics[width=\textwidth]{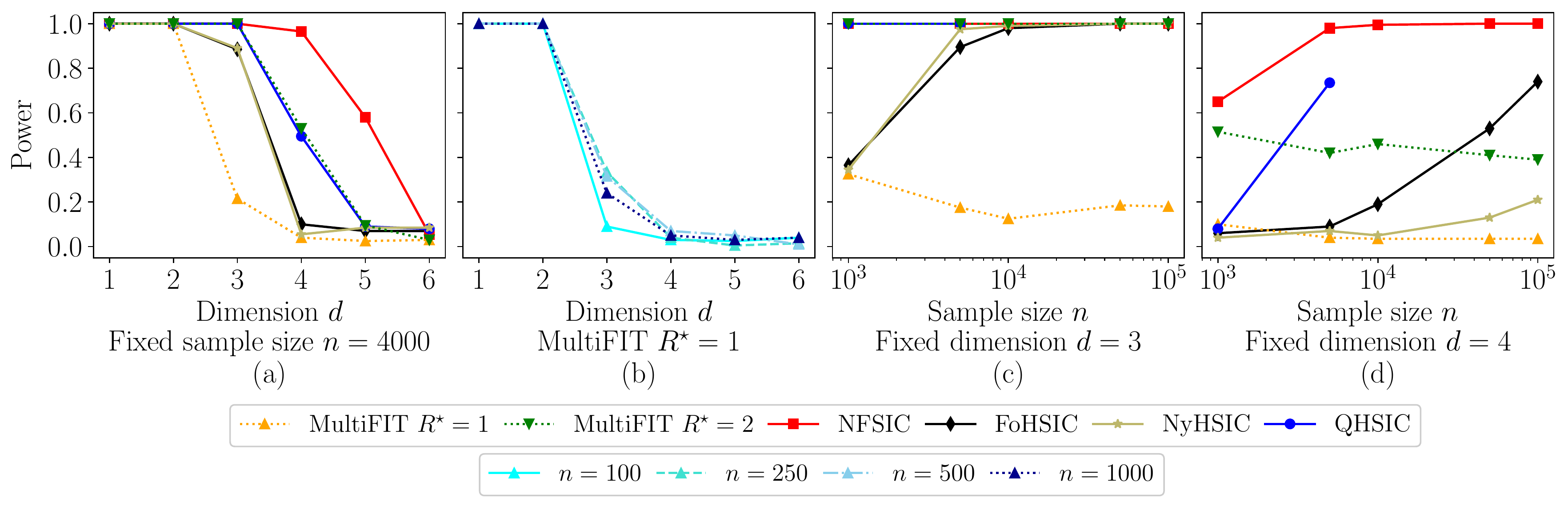}
\caption{Gaussian Sign experiment  where $X\sim\mathcal{N}(0,I_d)$ and $Y = |Z|\prod_{i=1}^d \mathrm{sgn}(X_i)$ with noise $Z\sim\mathcal{N}(0,1)$.}
\label{fig:gsign}
\end{figure}

In \Cref{fig:gsign}, for the Gaussian Sign experiment, the task is to detect the dependence between a $d$-dimensional Gaussian  $X\sim\mathcal{N}(0,I_d)$ and its noisy product of signs $Y = |Z|\prod_{i=1}^d \mathrm{sgn}(X_i)$ with noise $Z\sim\mathcal{N}(0,1)$, where $\mathrm{sgn}$ is the sign function.
In this setting, $Y$ depends on a combination  of all the features of $X$, but this dependence cannot be detected from any subset of the features of $X$.
For fixed sample size $n=4000$, we observe in \Cref{fig:gsign}~(a) that MultiFIT $R^\star=1$ suffers a significant loss of power from dimension 3 onwards, while all other tests retain high power. MultiFIT $R^\star=2$ performs better than NyHSIC and FoHSIC and as well as QHSIC, however NFSIC performs best, with power close to one even in dimension 4.
In \Cref{fig:gsign}~(b), we see that increasing the sample size for MultiFIT $R^\star=1$ does not necessarily increase the power.
This observation is confirmed in Figures~\ref{fig:gsign}~(c)~\&~(d) for $d=3$ and $d=4,$ where MultiFIT $R^\star=1$ has persistently low power.
In \Cref{fig:gsign}~(d), we see that the power of MultiFIT $R^\star=2$ also does not increase with sample size, and remains at approximately 0.5.
By contrast, the power of all  kernel tests increases with the sample size, as expected.

\begin{figure}[h]
\includegraphics[width=\textwidth]{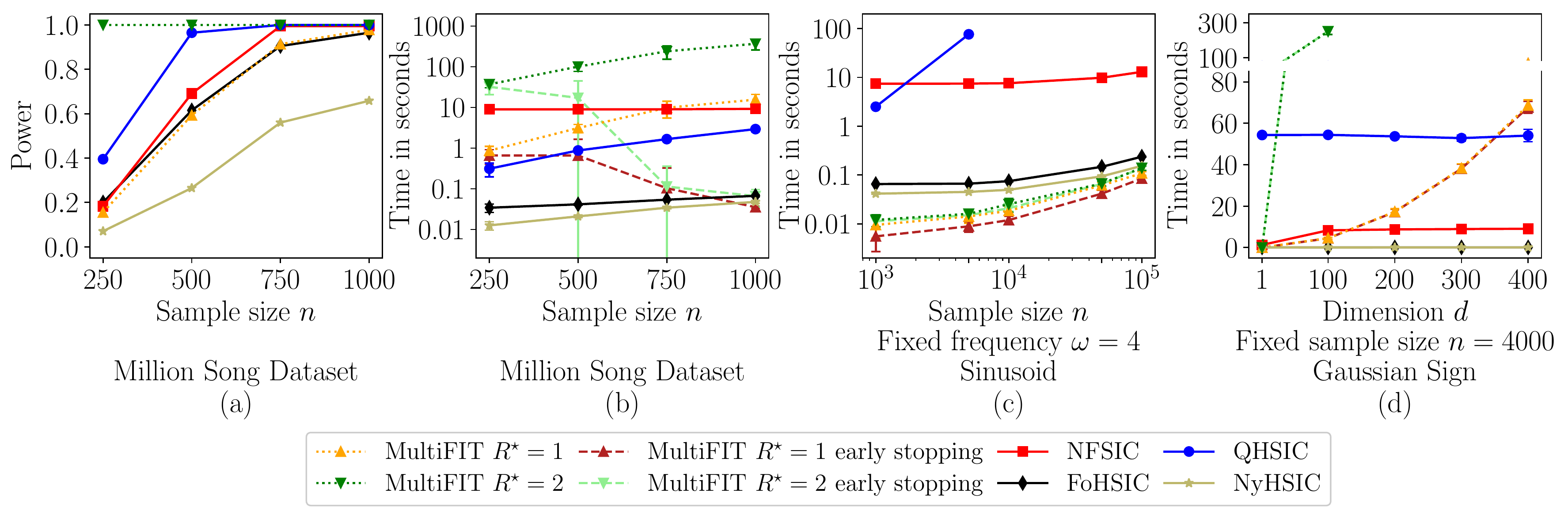}
\caption{
(a) Power experiment using the Million Song Dataset with $X$ consisting of 90 song features and $Y$ the song's release year. 
Running time experiments for the (b) Million Song Dataset, (c) Sinusoid, and (d) Gaussian Sign problems. 
}
\label{fig:msd}
\end{figure}

In \Cref{fig:msd}~(a), we use the Million Song Dataset \citep{bertin2011million} with the aim to capture the dependence between a song $X\in\mathbb{R}^{90}$ and its release year $Y\in\mathbb{N}\setminus\{0\}$. 
A song is represented by 90 standardized features consisting of 12 timbre averages and 78 timbre covariances.
In this real-world setting, we observe that MultiFIT $R^\star=1$ performs well: it outperforms NyHSIC and matches the power of FoHSIC which is only slightly below that of NFSIC. QHSIC performs even better, and MultiFIT $R^\star=2$ has power 1 for all considered sample sizes.
The strong performance of MultiFIT $R^\star=2$ comes at a computational cost, however, as observed in \Cref{fig:msd}~(b).
For sample size $n=250$, MultiFIT $R^\star=2$ takes almost two orders of magnitude more to run (in seconds) compared to MultiFIT $R^\star=1,$ and is much more expensive than all kernel tests.
As shown in \Cref{fig:msd}~(b), the computational complexity of MultiFIT in this experiment is much higher than $\mathcal{O}(n \log n)$, this corresponds to the setting described by \citet[Section 2.4]{gorsky2018multiscale} with alternatives which are `pervasive over the sample space and involve a large number of cuboids.'
As pointed out in their discussion, these large-scale global alternatives can often be detected in coarse resolutions, and the run times can be reduced significantly using early stopping, which is indeed what we observe.
It is interesting to see that, as the sample size increases, the problem becomes easier to solve, so MultiFIT can detect the dependence at coarser resolutions and stop the process early:  thus, MultiFIT tests with early stopping have computational times which decrease with the sample size. 
For the Sinusoid experiment (see \Cref{fig:msd}~(c)), MultiFIT runs the fastest and early stopping only slightly improves the run times; the computational times of the MultiFIT tests increase faster than those of the linear-time kernel tests.
As seen in \Cref{fig:msd}~(d) for the Gaussian Sign experiment, the computational complexity of MultiFIT (shown for $R^\star=1$) grows quadratically with the number of dimensions, while there is no notable increase in execution time for the kernel tests (whose only dimension-dependent cost  is in computing a dot product: linear with an extremely small constant). 
The run times of MultiFIT $R^\star=2$ explode with dimension: for $d=100$ it runs in roughly 4 minutes while all other linear-time tests run in less than 10 seconds.

\clearpage

\section*{Acknowledgements}
This article has been accepted for publication in Biometrika published by Oxford University Press.
It was supported by the Gatsby Charitable Foundation; and by the U.K.\ Research and Innovation [grant number EP/S021566/1].

\bibliography{paper-ref}

\begin{thebibliography}{}

\bibitem[Albert et~al., 2022]{albert2019adaptive}
Albert, M., Laurent, B., Marrel, A., and Meynaoui, A. (2022).
\newblock Adaptive test of independence based on {HSIC} measures.
\newblock {\em The Annals of Statistics}, 50(2):858--879.

\bibitem[Bertin-Mahieux et~al., 2011]{bertin2011million}
Bertin-Mahieux, T., Ellis, D.~P., Whitman, B., and Lamere, P. (2011).
\newblock The million song dataset.
\newblock {\em International Conference on Music Information Retrieval
  (ISMIR)}.

\bibitem[Chwialkowski and Gretton, 2014]{ChwGre14}
Chwialkowski, K. and Gretton, A. (2014).
\newblock A kernel independence test for random processes.
\newblock In {\em International Conference on Machine Learning (ICML)}, pages
  1422--1430.

\bibitem[Chwialkowski et~al., 2014]{ChwSejGre14}
Chwialkowski, K., Sejdinovic, D., and Gretton, A. (2014).
\newblock A wild bootstrap for degenerate kernel tests.
\newblock In {\em Advances in Neural Information Processing Systems (NIPS)},
  pages 3608--3616.

\bibitem[Gorsky and Ma, 2022]{gorsky2018multiscale}
Gorsky, S. and Ma, L. (2022).
\newblock Multiscale {F}isher's independence test for multivariate dependence.
\newblock {\em Biometrika}.

\bibitem[Gretton, 2015]{Gretton15}
Gretton, A. (2015).
\newblock A simpler condition for consistency of a kernel independence test.
\newblock Technical Report 1501.06103, ArXiv e-prints.

\bibitem[Gretton et~al., 2005]{GreBouSmoSch2005}
Gretton, A., Bousquet, O., Smola, A., and Sch{\"o}lkopf, B. (2005).
\newblock Measuring statistical dependence with {H}ilbert-{S}chmidt norms.
\newblock In {\em International Conference on Algorithmic Learning Theory
  (ALT)}, pages 63--77.

\bibitem[Gretton et~al., 2007]{GreFukTeoSonSchSmo08}
Gretton, A., Fukumizu, K., Teo, C., Song, L., Schoelkopf, B., and Smola, A.
  (2007).
\newblock A kernel statistical test of independence.
\newblock In {\em Advances in Neural Information Processing Systems (NIPS)},
  pages 585--592.

\bibitem[Gretton and Gy{\"o}rfi, 2010]{gretton2010consistent}
Gretton, A. and Gy{\"o}rfi, L. (2010).
\newblock Consistent nonparametric tests of independence.
\newblock {\em The Journal of Machine Learning Research}, 11:1391--1423.

\bibitem[Jitkrittum et~al., 2017]{jitkrittum2017adaptive}
Jitkrittum, W., Szab{\'{o}}, Z., and Gretton, A. (2017).
\newblock An adaptive test of independence with analytic kernel embeddings.
\newblock In {\em International Conference on Machine Learning (ICML)}, pages
  1742--1751.

\bibitem[Kim et~al., 2022]{kim2020minimax}
Kim, I., Balakrishnan, S., and Wasserman, L. (2022).
\newblock Minimax optimality of permutation tests.
\newblock {\em The Annals of Statistics}, 50(1):225--251.

\bibitem[Li and Yuan, 2019]{li2019optimality}
Li, T. and Yuan, M. (2019).
\newblock On the optimality of {G}aussian kernel based nonparametric tests
  against smooth alternatives.
\newblock {\em arXiv preprint arXiv:1909.03302}.

\bibitem[Lyons, 2013]{lyons2013}
Lyons, R. (2013).
\newblock {Distance covariance in metric spaces}.
\newblock {\em The Annals of Probability}, 41(5):3284--3305.

\bibitem[Rahimi and Recht, 2007]{rahimi07random}
Rahimi, A. and Recht, B. (2007).
\newblock Random features for large-scale kernel machines.
\newblock In {\em Advances in Neural Information Processing Systems (NIPS)},
  pages 1177--1184.

\bibitem[Schrab et~al., 2022]{schrab2022efficient}
Schrab, A., Kim, I., Guedj, B., and Gretton, A. (2022).
\newblock Efficient aggregated kernel tests using incomplete {$U$}-statistics.
\newblock {\em arXiv preprint arXiv:2206.09194}.

\bibitem[Sejdinovic et~al., 2013]{SejSriGreFuk2013}
Sejdinovic, D., Sriperumbudur, B., Gretton, A., and Fukumizu, K. (2013).
\newblock {{Equivalence of distance-based and RKHS-based statistics in
  hypothesis testing}}.
\newblock {\em Annals of Statistics}, 41(5):2263--2291.

\bibitem[Sriperumbudur et~al., 2010]{SriGreFukSchetal10}
Sriperumbudur, B., Gretton, A., Fukumizu, K., Sch{\"o}lkopf, B., and Lanckriet,
  G. (2010).
\newblock Hilbert space embeddings and metrics on probability measures.
\newblock {\em Journal of Machine Learning Research}, 11:1517--1561.

\bibitem[Szab{{\'o}} and Sriperumbudur, 2018]{SzaSri18}
Szab{{\'o}}, Z. and Sriperumbudur, B. (2018).
\newblock Characteristic and universal tensor product kernels.
\newblock {\em Journal of Machine Learning Research}, 18(233):1--29.

\bibitem[Sz{\'e}kely and Rizzo, 2009]{szekely2009}
Sz{\'e}kely, G.~J. and Rizzo, M.~L. (2009).
\newblock {Brownian distance covariance}.
\newblock {\em The Annals of Applied Statistics}, 3(4):1236--1265.

\bibitem[Williams and Seeger, 2001]{williams01using}
Williams, C. and Seeger, M. (2001).
\newblock Using the {N}ystr{\"{o}}m method to speed up kernel machines.
\newblock In {\em Advances in Neural Information Processing Systems (NIPS)},
  pages 682--688.

\bibitem[Zhang et~al., 2018]{zhang2018large}
Zhang, Q., Filippi, S., Gretton, A., and Sejdinovic, D. (2018).
\newblock Large-scale kernel methods for independence testing.
\newblock {\em Statistics and Computing}, 28(1):113--130.

\end{thebibliography}

\end{document}